\documentclass[12pt]{article}
\usepackage[dvips]{graphicx}
\usepackage{epsfig}
\usepackage{lineno}
\usepackage{color}
\usepackage{epstopdf}

\newcommand{\grl}{    {Geophys. Res. Lett.}}
\newcommand{\jgr}{    {J. Geophys. Res.}}

\begin{document}
\begin{center}
{\Large \bf Extreme time-integrated geomagnetic activity: ap index statistics}\\

\vskip 10mm

{\Large D. Mourenas $^{1}$, A. V. Artemyev $^{2}$, X.-J. Zhang $^{2}$,\\ V. Angelopoulos $^{2}$}\\

\vskip 5mm

{\large \it $^{1}$ CEA, DAM, DIF, Arpajon, France}\\
{\large \it $^{2}$ Institute of Geophysics and Planetary Physics, University of California, Los Angeles, USA}\\

\end{center}

\begin{abstract}
We analyze statistically extreme time-integrated $ap$ events in 1958-2007, which occurred during both strong and weak geomagnetic storms. The tail of the distribution of such events can be accurately fitted by a power-law with a sharp upper cutoff, in close agreement with a second fit inferred from Extreme Value Theory. Such a behavior is suggestive of a self-organization of the solar wind-magnetosphere-ionosphere system appearing during strong and sustained solar wind driving. The 1 in 10 years to 1 in 100 years return levels of such extreme events are calculated, taking into account possible solar cycle modulations. The huge October 2003 event turns out to be a 1 in 100$\pm40$ years event. Comparisons with the distribution of extreme time-integrated $aa$ events collected in 1870-2010 support the reliability of our results over the long run. {Using data from Time History of Events and Macroscale Interactions during Substorms (THEMIS) satellites and the Van Allen Probes, we show that extreme time-integrated $ap$ events produce hard fluxes of energetic electrons and ions in the magnetotail and high fluxes ($>10^6$ e/cm$^2$/sr/s/MeV) of 1.8 MeV electrons in the heart of the outer radiation belt. }
\end{abstract}

\section{Introduction}

Space weather is known to pose an important threat for satellites, due to the high fluence of energetic and relativistic electrons produced during disturbed periods \cite{HorneSW13, Ferguson2015, Ganushkina2017, Iucci2005, Schrijver2015}. Many studies have attempted to determine the most pertinent geomagnetic indices for characterizing such periods of high risk. $Dst$ index \cite{Mayaud80} minima, corresponding to peak storm disturbance, show some correlation with MeV electron flux enhancements \cite{Reeves2003} and satellite anomalies when $Dst<-70$ nT \cite{Iucci2005}. They are well correlated with maxima of relativistic electron phase space density at $L=4-5$ during periods of simultaneously elevated $AE$ \cite{Tang17, Zhao17}. High levels of $IntDst$, the time-integrated $Dst$ when $Dst<-70$ nT (a limit corresponding to strong chorus-induced electron energization, see \cite{Artemyev15:nature}), are correlated with 2-4 MeV electron flux maxima at $L=4.2$, high time-integrated $AE$, and harder spectra of particle injections from the magnetotail \cite{Mourenas18}.

However, \cite{Wrenn1996} and \cite{Iucci2005} have noticed that periods of high $Kp>3^+$ or $ap>21$ also correspond to high levels of satellites anomalies, while \cite{Wrenn1987} emphasized the importance of considering a time-integrated $ap$ index for studying adverse effects of space weather on satellites. The $ap$ index (varying linearly from 0 to 400) and the $Kp$ index (varying quasi-logarithmically from 0 to 9) are equivalent geomagnetic indices obtained from middle latitude ($\sim 44^\circ-60^\circ$) 3-hour measurements of the disturbance range of the horizontal component of the Earth's magnetic field, whereas the hourly $Dst$ index is derived from low latitude measurements of the average horizontal component of the geomagnetic field (e.g., \cite{Mayaud80, Lockwood2013, Ganushkina2017}). For the sake of simplicity, the $ap$ index will henceforth be written in units of nT as $Dst$, although the average unit of the $ap$ index is actually closer to $\approx 2$ nT (e.g., \cite{Lockwood2018}). The $Dst$ index provides an estimate of the strength of the storm time ring current, whereas the $ap$ index provides a measure of magnetospheric convection and substorm currents that bring inside geosynchronous orbit $\sim 5-300$ keV electrons as well as energetic ions \cite{Borovsky17b, Lockwood2018, Thomsen2013}.

As a result, the two main contributors to high fluxes of 'satellite-killer' MeV electrons in the outer radiation belt -- electron inward transport via convection or radial diffusion by Ultra Low Frequency (ULF) waves, and local energization of 100 to 300 keV electrons by chorus waves excited by 5 to 30 keV electrons -- are better correlated with $ap$ than $Dst$ (e.g., see \cite{Borovsky17, Borovsky17b, Boyd18, Horne18, Meredith2003, Ozeke14}). Important differences between $ap$ and $Dst$ indices most notably show up during high-speed solar wind streams that trigger substorms and weak magnetic storms. Such disturbances often produce elevated fluxes of MeV electrons at geosynchronous orbit, leading to significant internal charging and radiation dose effects in satellites, but barely affect low latitude magnetometers that provide the $Dst$ index (e.g., \cite{Borovsky17, Ganushkina2017, Horne18}). MeV electron fluxes at $L=4-6$ measured on low Earth orbit are better correlated with periods of $ap>22$ nT than with $Dst$ \cite{Katsiyannis2018}. Satellite surface charging hazards, related to intense fluxes of 1-50 keV electrons injected from the plasma sheet, are similarly better correlated with $Kp$ than $Dst$ \cite{Wrenn1996, Thomsen2013}. Geomagnetically induced currents caused by relatively high latitude currents should also better correlate with $ap$ than $Dst$ \cite{Borovsky17b}.

Moreover, \cite{Iucci2005} have pointed out (see their Figure 9a) that geosynchronous satellite anomalies generally occur near the end of $\sim$ 4-day periods of mean daily $ap>21$ nT and time-integrated $ap$ $>2100$ nT$\cdot$hr. Similarly, \cite{Ferguson2015} found a correlation between satellite anomalies and days with a sum of $Kp$ values larger than 35. At geostationary orbit, \cite{Kim2015} found a mean $Kp>3$ during strong enhancements of 2 MeV electron flux, and \cite{Borovsky17} obtained a significant correlation between 1.2 MeV electron flux and $Kp$ integrated in time over a fixed period of tens of hours. Therefore, it is crucial for space weather studies to supplement our earlier analysis of the time-integrated $Dst$ index \cite{Mourenas18} by a similar statistics of the time-integrated $ap$ index, hereafter denoted $Int(ap)$.

In the next section, we shall provide a statistics of extreme $Int(ap)$ events, as well as best fits to their distribution. We shall examine their properties, comparing them to time-integrated $Dst$ and $aa$ events, showing their correlations with {solar wind-magnetosphere coupling functions} and sunspot number, and providing estimates of their return levels. {We shall further propose a simple predictor of the strength of extreme $Int(ap)$ events, based on parameters obtained during the early phase of these events. Finally, we will briefly examine the relationships between $Int(ap)$, energetic particle injections in the magnetotail, and relativistic electron flux in the heart of the outer radiation belt.}

\section{Extreme time-integrated $ap$ statistics}

\subsection{Data set of extreme events}

{In this study, we make use of a 1958-2017 dataset of the $ap$ index obtained from the World Data Center in Kyoto. We shall first investigate the probability distribution of extreme $In(ap)$ events over the 1958-2007 period, to allow meaningful intercomparisons with previous statistics of time-integrated $IntDst>670$ nT$\cdot$hr events \cite{Mourenas18} that were based on the 1958-2007 dataset of revised $Dst$ index provided by \cite{Love2009}. Next, we shall consider the 2000-2017 period to explore possible correlations between $Int(ap)$ and various solar wind parameters, because this 2000-2017 period is characterized by a relatively high availability of solar wind data in the OMNI data base, especially as compared to previous times..}

{First, all extreme time-integrated events with $Int(ap)>975$ nT$\cdot$hr have been compiled over 1958-2007.} An integration threshold $ap\geq 22$ nT (equivalent to $Kp>3^+$) has been used for the time-integrated $Int(ap)$ events. This particular threshold was chosen based on applied studies of space weather, as well as on physical grounds. Statistics of satellite anomalies indeed suggest that $Kp>3^+$ or $ap>21$ nT periods correspond to significantly higher risks \cite{Wrenn1996, Iucci2005, Ferguson2015}. Moreover, chorus-induced energization of relativistic electrons in the outer radiation belt increases abruptly and significantly between $Kp=3$ and $Kp=4$ based on a recent synthetic chorus wave model derived from Cluster and Van Allen Probes statistics, reaching efficient energization rates $D_{EE}/E^2\approx 1$ day$^{-1}$ (see Figure 14 from \cite{Agapitov18}). Another important contribution to relativistic electron acceleration comes from their inward radial diffusion by ULF waves \cite{Ozeke14, Zong2017}. Radial diffusion rates also increase significantly between $Kp=3$ and $Kp=4$, reaching efficient levels $D_{LL}\approx 1$ day$^{-1}$ at $L\sim 6$ \cite{Ozeke14, Murphy2016}, further justifying the use of some threshold $Kp>3+$ or $ap>22$ nT. {Each $Int(ap)$ event is selected through a peak-over-threshold method as the period comprised between a first time when $ap$ increases above the $ap=22$ nT threshold and the next time when $ap$ decreases below this same $ap=22$ nT threshold, ensuring the independence of $Int(ap)$ events, which are separated by periods of low activity $ap<22$ nT (e.g., see \cite{Coles2001, Mourenas18, Tsubouchi2007}).}

\begin{figure*}
\includegraphics[width=36pc]{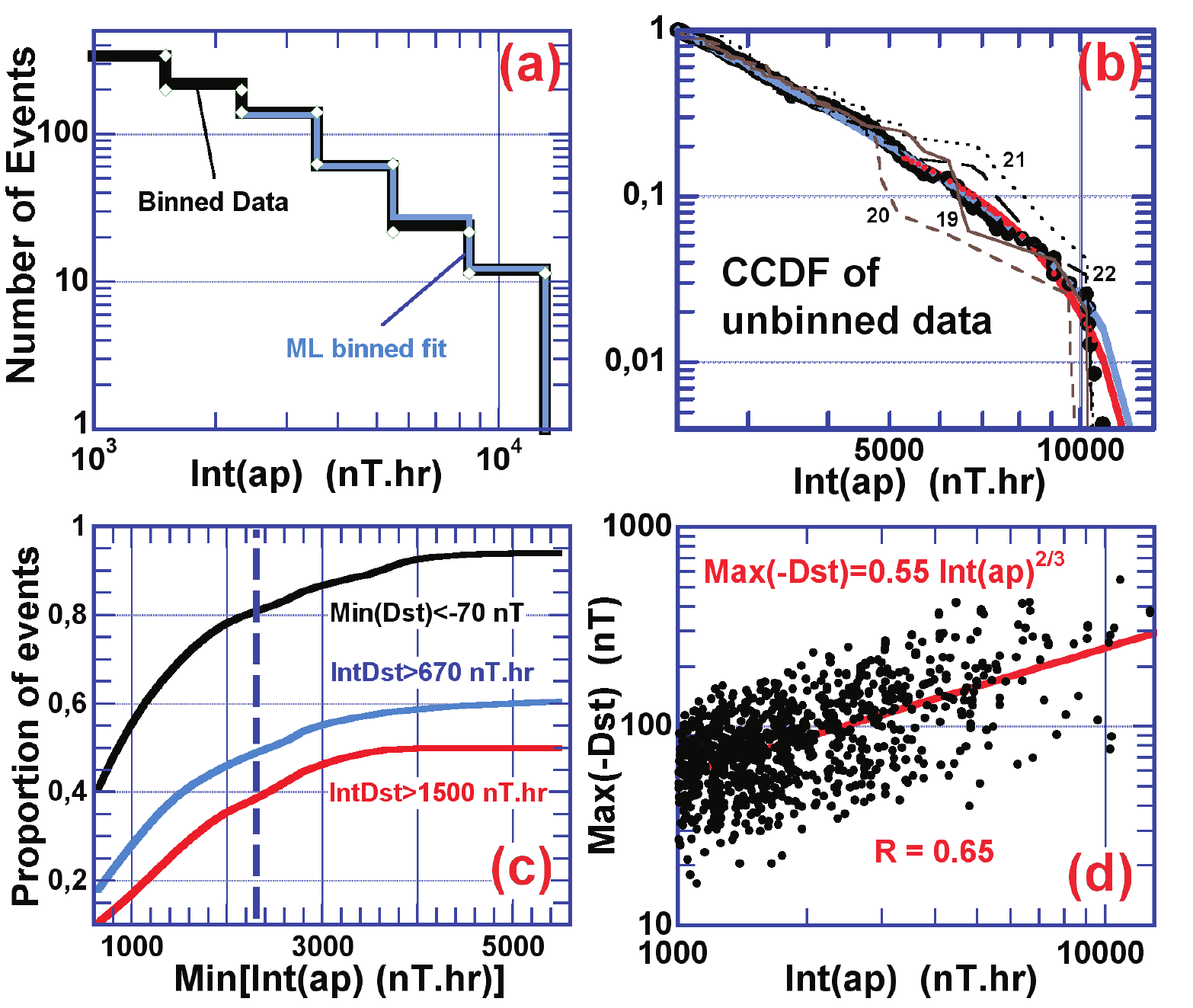}
\end{figure*}
\begin{figure*}
\caption{(a) Fifty-year (1958-2007) binned distribution of the number of intense $Int(ap)$ events with $ap\geq 22$ nT and $Int(ap)>975$ nT$\cdot$hr (solid black line) {and Maximum Likelihood power-law fit (solid blue line).} Binned data obtained for a threshold $ap\geq 27$ nT instead of $ap\geq 22$ nT (with $Int(ap)$ renormalized by a factor 1.14) are also displayed (white lozenges). (b) CCDF of 1958-2007 unbinned data (black circles) with $Int(ap)>2310$ nT$\cdot$hr with corresponding ML fit in the form of a power-law with upper-cutoff (blue). Another ML fit in the form of a generalized Pareto distribution is shown in red for $Int(ap)>5200$ nT$\cdot$hr. The CCDF of unbinned data during four different solar cycles is indicated by thin black curves. (c) Proportion of $Int(ap)>\min[Int(ap)]$ events in 1958-2007 with corresponding minimum $Dst<-70$ nT (black), $IntDst>670$ nT$\cdot$hr (blue), and $IntDst>1500$ nT$\cdot$hr (red), as a function of $\min[Int(ap)]$. A vertical dashed line indicates 2310 nT$\cdot$hr. (d) Max$(-Dst)$ versus $Int(ap)$ for events in 1958-2017 with corresponding power-law fit (red line). }
\label{fig01}
\end{figure*}

The binned distribution of our events is shown in Figure \ref{fig01}a. We used 6 logarithmically-spaced bins with a factor 1.54 between minimum and maximum bin limits, to ensure that there was a sufficient number of events (at least 10) in each bin for an accurate fitting procedure \cite{Press1992}. Among these 798 events, we found that 55.5 \% occurred during storms with minimum $Dst<-70$ nT, while 28 \% had also a time-integrated $Dst$ index $IntDst>670$ nT$\cdot$hr (integrated in time over the same periods as $Int(ap)$ events as long as $Dst<-70$ nT), and 17 \% had $IntDst>1500$ nT$\cdot$hr. Consistent with this, $\sim 60$ \% of all 1958-2007 events with $IntDst>670$ nT$\cdot$hr (simply integrated in time as long as $Dst<-70$ nT) found in a previous study \cite{Mourenas18} had also $Int(ap)>975$ nT$\cdot$hr. Thus, many events are simultaneously extreme $Int(ap)$ and extreme $IntDst$ events. Figure \ref{fig01}c shows that the proportion of such doubly extreme events increases with $Int(ap)$, reaching $\approx 50$ \% for $Int(ap)>2310$ nT$\cdot$hr. This likely corresponds to a deeper and more sustained penetration of energetic particles injected from the magnetotail during stronger $Int(ap)$ events, generally producing a stronger peak Max($-Dst$) disturbance (see Figure \ref{fig01}d) as well as stronger $IntDst$ events.

Despite some global correlation between extreme $Int(ap)$ and $IntDst$, however, extreme $Int(ap)$ events do not always correspond to extreme $IntDst$ levels. {For instance, among the 8 strongest $Int(ap)$ events of 1958-2007, the important 4 August 1972 storm \cite{Knipp18} reached $Int(ap)=9580$ nT$\cdot$hr but attained only much weaker $IntDst=1650$ nT$\cdot$hr and $\min(Dst)\sim-110$ nT levels than the March 1989 ($IntDst=10300$ nT$\cdot$hr, $\min(Dst)\sim-550$ nT) and October 2003 ($IntDst=9510$ nT$\cdot$hr, $\min(Dst)\sim-372$ nT) superstorms.} Special conditions are probably needed for a sufficiently deep and prolonged penetration of injected particles in the inner magnetosphere, so that they can generate a ring current sufficiently strong to produce durable low latitude magnetometer signatures resulting in a high $IntDst$. Such conditions do not depend exclusively on the sole $ap$ index.

Hereafter, we shall mainly focus on the 236 most important events, with $Int(ap)>2310$ nT$\cdot$hr, composing the tail of the distribution displayed in Figure \ref{fig01}a. The threshold was fixed at 2310 nT$\cdot$hr for the following reasons: (i) it corresponds to the approximate threshold above which $Int(ap)$ events are more likely to be also strong $IntDst$ events in Figure \ref{fig01}c, (ii) this part of the distribution is the most stable when the integration threshold is varied (increasing it from $ap\geq 22$ nT to $ap\geq 27$ nT leads to a similar distribution shape at $Int(ap)>2310$ nT$\cdot$hr after renormalization in Figure \ref{fig01}a), (iii) the slope of the distribution decreases at $Int(ap)<2310$ nT$\cdot$hr (the distribution shape becomes less meaningful at lower $Int(ap)$ because the $ap\geq 22$ nT threshold then leads to a saturation of the number of events), and (iv) \cite{Iucci2005} have shown that geosynchronous satellite anomalies generally occur near the end of periods with $Int(ap)>2100$ nT$\cdot$hr.

\subsection{Best fits to the tail of the $Int(ap)$ distribution and physical interpretation}

When examining the binned distribution of $Int(ap)$ events in Figure \ref{fig01}a, the empirical data at $x=Int(ap)> x_{min}=2310$ nT$\cdot$hr seems to follow a power-law yearly probability distribution of the form $P_y[x]= C\cdot{\rm H}(x_{max}-x)/x^{\alpha}$, with ${\rm H}$ the Heaviside function and without any event above an upper cutoff $x_{max}\simeq 13000$ nT$\cdot$hr {-- a form similar to the distribution of extreme time-integrated $IntDst$ events (see \cite{Mourenas18}).}

{To check the pertinence of such a power-law shape, we calculated a maximum likelihood (ML) fit of the above form $P_y$ to the observed distribution of $Int(ap)>2310$ nT$\cdot$hr events in Figure \ref{fig01}a,b (for details on these techniques, see, e.g., \cite{Clauset2009, Love2015, Mourenas18, Press1992}), giving $\alpha =2.85$ and $C=1.52\cdot10^7$. The delta method applied to the ML estimate (e.g., \cite{Coles2001}) gives a 95\% confidence interval $2.53<\alpha<3.17$. The goodness of this fit (displayed in Figure \ref{fig01}a,b with the observed distribution) has been checked using the Kolmogorov-Smirnov (KS) test \cite{Clauset2009, Press1992}. The maximum KS distance $D\simeq 0.042$ to the complementary to the cumulative distribution function (CCDF) of unbinned data above 2310 nT$\cdot$hr corresponds to a $p$-value of $p=0.66$. The ML method therefore provides a plausible fit to the data, indicating that the power-law distribution hypothesis cannot be confidently rejected.}

\begin{figure*}
\includegraphics[width=36pc]{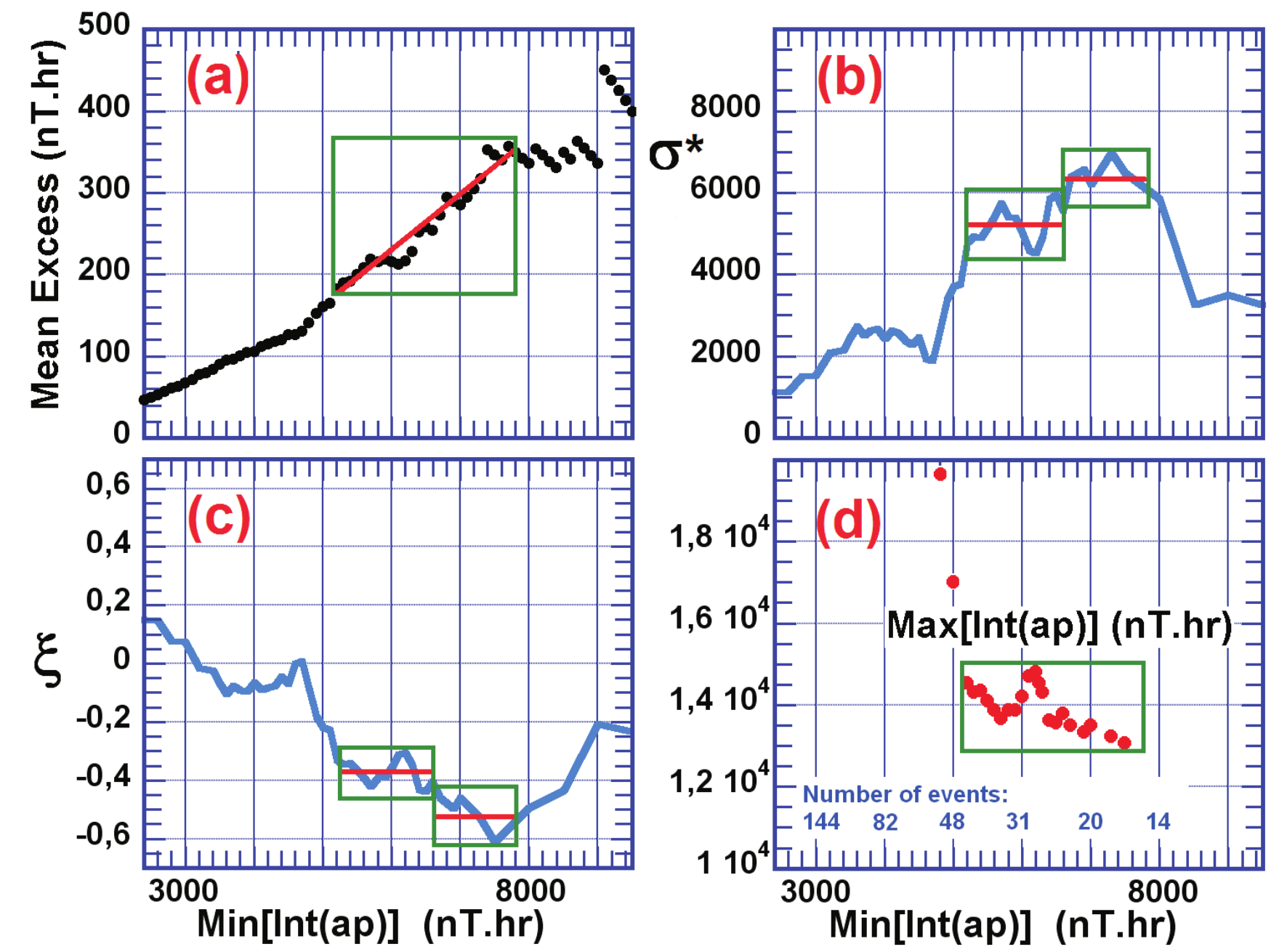}
\end{figure*}
\begin{figure*}
\caption{(a) Mean excess $\langle Int(ap)-\min[Int(ap)]\rangle$ as a function of threshold $\min[Int(ap)]$ for intense $Int(ap)$ events with $ap\geq 22$ nT in 1958-2007. The red curve shows a linear fit. (b) Rescaled $\sigma^*=\sigma -\xi\cdot\min[Int(ap)]$ parameter of the Generalized Pareto Distribution fit, estimated by Maximum Likelihood as a function of threshold $\min[Int(ap)]$. (c) $\xi$ parameter of the ML GPD fit as a function of threshold $\min[Int(ap)]$. (d) Extreme Value Theory estimates of $\max[Int(ap)]$, inferred from ($\sigma, \xi$) GPD parameters as a function of $\min[Int(ap)]$ (with corresponding number of events above threshold indicated at the bottom).}
\label{fig02}
\end{figure*}

{Another way to model the probability of extreme $Int(ap)$ events is to use} Extreme Value Theory (EVT), which was especially designed to study such stochastic rare events {\cite{Coles2001, Meredith14, Tsubouchi2007}.} In analogy to the central limit theorem, it has been shown that the exceedances over a threshold in a sample of $N$ independent events tend to follow a Generalized Pareto Distribution (GPD) for sufficiently high $N$ and threshold values \cite{Coles2001}. A reliable EVT method therefore consists in fitting the tail of the distribution of exceedances {of independent (by construction) $Int(ap)$ events} over a well-chosen and sufficiently high threshold $\min[Int(ap)]$ by a GPD of the form $P_{GPD}(\xi,\sigma)=(1+\xi\cdot[Int(ap) -\min[Int(ap)]]/\sigma)^{-1/\xi-1}/\sigma$ for its probability distribution (e.g., see \cite{Coles2001} and references therein).

The convenient $\min[Int(ap)]$ threshold domain for a reliable GPD fit can be determined by different complementary techniques. It should correspond to (i) sufficiently high $\min[Int(ap)]$ threshold values in the distribution tail (typically within the $\sim 30-60$ upper data points), (ii) a roughly linear relationship between mean excess $\langle Int(ap)-\min[Int(ap)]\rangle$ and threshold $\min[Int(ap)]$, and (iii) nearly constant (stable) estimates of GPD parameters $\xi$ and $\sigma^*=\sigma -\xi\cdot\min[Int(ap)]$ obtained via Maximum Likelihood \cite{Coles2001,Tsubouchi2007}. In Figure \ref{fig02}a, the mean excess (black points) first increases logarithmically up to 4800 nT$\cdot$hr, then it has a sudden inflexion, before showing some evidence of linearity over the region $\min[Int(ap)]\sim 5200-7800$ nT$\cdot$hr (see red line) corresponding to the 50 upper data points. Inside the same $\min[Int(ap)]$ region, $\xi$ and $\sigma^*$ are simultaneously nearly constant over two $\min[Int(ap)]$ domains shown by green boxes in Figures \ref{fig02}b,c.

The optimal $(\xi,\sigma)$ parameters generally correspond to the lowest threshold inside the above-determined convenient threshold domain, because more events are then taken into account \cite{Coles2001}. {In the present case, this led us to use the lowest convenient threshold $\min[Int(ap)]=5200$ nT$\cdot$hr, giving a shape parameter $\xi\simeq -0.33 \pm 0.29$ and a scale parameter $\sigma\simeq 3060 \pm 1250$, with minimum and maximum parameter values corresponding to 95\% confidence intervals calculated via the delta method \cite{Coles2001}. Considering a slightly higher threshold $\min[Int(ap)]=5500$ nT$\cdot$hr (corresponding to the mean $\xi$ and $\sigma$ shown by red lines in Figure \ref{fig02}b,c) would have given very similar $\xi\sim -0.36 \pm 0.32$ and $\sigma\sim 3200 \pm 1400$ values but with larger uncertainty, leading us to hereafter keep the first values in accordance with usual practice \cite{Coles2001}.} This GPD fit, shown in Figure \ref{fig01}b (red curve), is fairly close to both the power-law fit and the data, with a maximum KS distance $D\simeq 0.1$ corresponding to $p=0.66$.

Moreover, Extreme Value Theory confirms the likely presence of an upper limit $\max[Int(ap)]\simeq 13000-14500$ nT$\cdot$hr (see Figure \ref{fig02}d) very close to the upper cutoff assumed for the power-law fit, and only slightly larger than the strongest $Int(ap)=12700$ nT$\cdot$hr October 2003 event encountered during the 1958-2007 period. {Since both the shape $\xi$ and scale $\sigma$ parameters and the corresponding $\max[Int(ap)]$ of GPD fits vary abruptly and strongly below $Int(ap)=5200$ nT$\cdot$hr, however, no unique GPD fit can be used over the whole domain $Int(ap)>2310$ nT$\cdot$hr of the distribution tail, which is more conveniently described by the ML power-law fit.}

{Physically, such power-law distributions of the most extreme $Int(ap)$ and $IntDst$ events could result from protracted periods of strong solar wind driving that compel the magnetosphere to assume a particular self-organized critical system configuration in nearly stable non-equilibrium (e.g., see \cite{Angelopoulos1999, Aschwanden16, Uritsky2001, Valdivia2013, Welling2016}).} When considering less disturbed periods at $Int(ap)<1500$ nT$\cdot$hr, a lognormal distribution would be probably more relevant (see Figure \ref{fig01}a), as for the full $Dst$ index distribution \cite{Love2015}. A lognormal distribution of event sizes can easily be produced by independent multiplicative random processes, when each multiplicative process has finite mean and variance and follows Gibrat's law of proportionate effect (such that the relative growth of an event is independent of its size, see \cite{Mitzenmacher03, Montroll82} and references therein). The lognormality of the $Dst$ index up to high levels \cite{Love2015}, as well as the lognormality of $Int(ap)$ below $2300$ nT$\cdot$hr in Figure \ref{fig01}a, are probably related to various independent multiplicative random factors that affect the strength of geomagnetic disturbances \cite{Love2015, Lockwood2018}.

{For the largest $Int(ap)$ events, however, sizes instead follow a power-law (also called Pareto) distribution (see \cite{Montroll82} on the analogous behavior of the size of incomes in economy).} What can explain this evolution from lognormal to power-law? Actually, it is well-known that a small change in the lognormal generative process can produce a power-law distribution \cite{Mitzenmacher03}. {The largest $Int(ap)$ geomagnetic events may become partly self-sustaining (e.g., \cite{Klimas2000, Liu2013, Valdivia2013}), which would allow more amplification over longer periods. The presence of an additional amplification for stronger events can produce power-law distributions with exponents $1\leq\alpha\leq 3$ \cite{Montroll82}.} The neat power-law distribution with sharp upper cutoff found in Figure \ref{fig01} further suggests a saturation just below 13000 nT$\cdot$hr, consistent with the GPD fit. This development of a saturation process at very high $Int(ap)$ could simply represent the last stage in the self-organization of the solar wind-magnetosphere-ionosphere system that already accounts for the power-law distribution (e.g., \cite{Aschwanden16, Valdivia2013, Welling2016}).

Nevertheless, there is another possible scenario. The saturation process taking place below $Int(ap)=13000$ nT$\cdot$hr may be viewed as an exponentially increasing drag on growth as size increases. Thus, this saturation process can lead to a slower decrease of the originally lognormal distribution of grown events near the upper limit, potentially leading to an approximate power-law. In fact, the effect of such a fast increasing drag near the upper bound is roughly similar to the effect of an upper reflecting barrier. There is also a lower reflecting barrier at very small $Int(ap)=66$ nT$\cdot$hr, since newborn $Int(ap)$ events cannot decrease below this level by construction. When independent random multiplicative processes following Gibrat's law of proportionate effect are operating between lower and upper reflecting barriers, the asymptotic stable distribution is often a power-law distribution \cite{Zhang10}. In this second scenario, the only form of self-organization of the solar wind-magnetosphere-ionosphere system would be the saturation process itself.

\subsection{{Relationships of $Int(ap)$ with sunspot number and solar wind-magnetosphere coupling parameters}}

\begin{figure*}
\includegraphics[width=36pc]{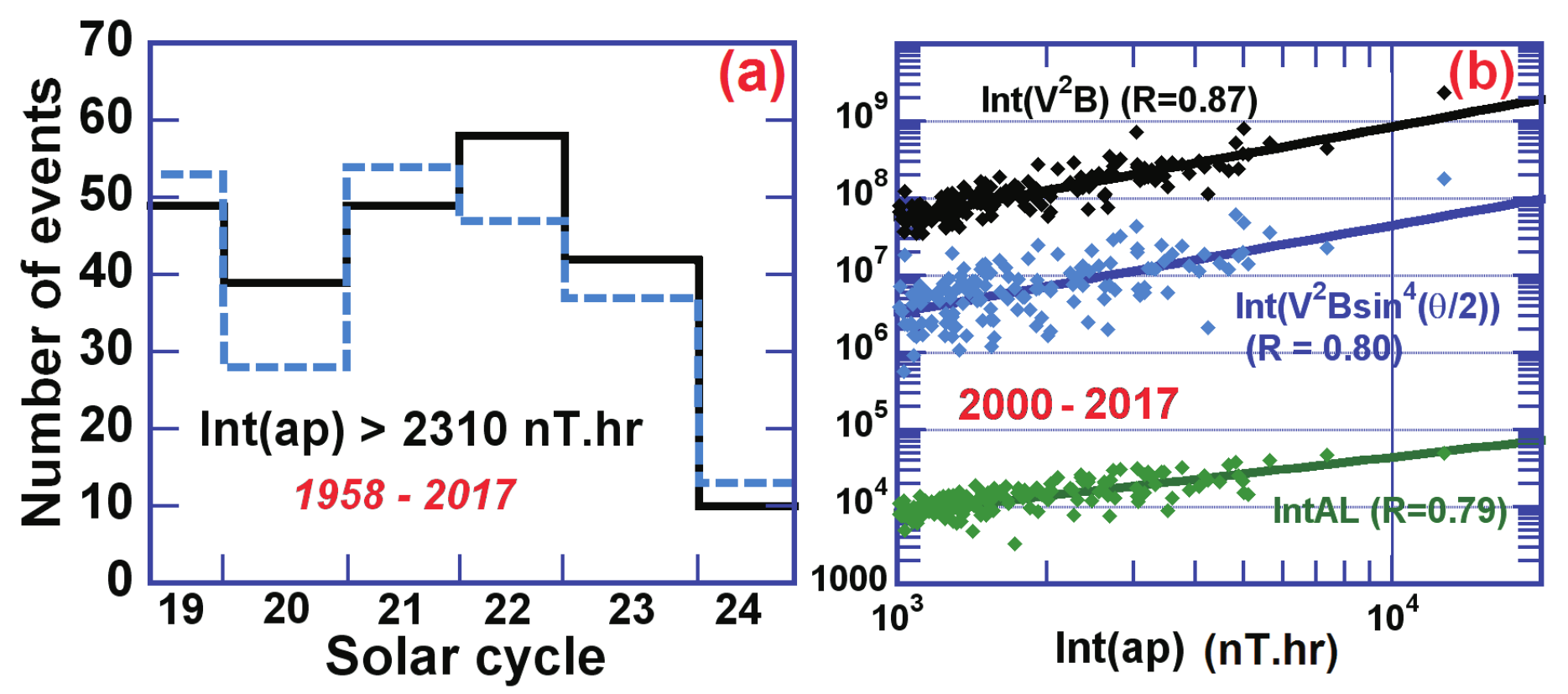}
\caption{(a) Number of strong $Int(ap)>2310$ nT$\cdot$hr events during each solar cycle (or partial cycle) in 1958-2017 estimated by the probability $P_y(\alpha=2.85)\times S_C$ (dashed blue lines) versus observed occurrences (black). {(b) Correlations between $Int(ap)$ and $Int(V_{sw}^2B)$, $Int(V_{sw}^2B\sin^4(\theta/2))$, and $IntAL$ during significant $Int(ap)>1000$ nT$\cdot$hr events in 2000-2017 (black, blue, and green points, respectively), together with corresponding power law least squares fits (solid lines of the same colors).} }
\label{fig03}
\end{figure*}

Figure \ref{fig01}b shows that the complementary to the cumulative distribution function (CCDF) of extreme {$Int(ap)>2310$ nT$\cdot$hr} events remains similar from one solar cycle to another. {However, Figure \ref{fig03}a demonstrates that the total number of $Int(ap)$ events does vary from one solar cycle to the other (black line). Interestingly, the number of $Int(ap)$ events during each solar cycle has roughly a modulation $\sim S_C=(\max[S_{n,y}]/210)^2$ with the maximum yearly mean sunspot number $\max[S_{n,y}]$ during each cycle (dashed blue line), similar to the modulation of the number of extreme $IntDst$ events per cycle \cite{Mourenas18}. The maximum value of the yearly mean sunspot number $S_{n,y}$ during a solar cycle is an approximate measure of solar activity during a given cycle. The above formulation of $S_C$ was derived by \cite{Mourenas18} to satisfy the following 3 requirements: (i) approximately reproducing the modulation of the probability of extreme $IntDst$ events with solar cycle in 1958-2007, (ii) not altering the global 1958-2007 probability distribution $P_y$ of $IntDst$ when multiplying it by $S_C$, and (iii) using the simplest analytical form. The obtained simple formulation of $S_C$ \cite{Mourenas18} is indeed such that $\langle\max[S_{n,y}]\rangle=1$ over solar cycles 19-23. The normalization factor 210 of $\max[S_{n,y}]$ also nearly corresponds to the average value ($=205$) of $\max[S_{n,y}]$ over solar cycles 19 to 23. Such a simple formulation allows the multiplicative factor $S_C$ to approximately reproduce the modulation of the probability of both extreme $Int(ap)$ and $IntDst$ events with solar cycle, while not altering their global 1958-2007 statistics \cite{Mourenas18}. A better fitting formula might possibly be obtained by considering the fine details of the distribution of sunspot number during each cycle, but this is beyond the scope of the present paper. A correlation of the $aa$ index with sunspot number has also been noticed previously by \cite{Love2011}. }

{Next, we examined possible correlations between significant $Int(ap)>1000$ nT$\cdot$hr and time-integrated solar wind parameters in 2000-2017 -- a period of relatively high availability of solar wind data.} Solar wind velocity $V_{sw}$ values on 29-30 October 2003 which were missing in the OMNI data base were replaced by actual data from \cite{Skoug2004}. Missing $V_{sw}$ data during 3 other periods in 2001 and 2005 were replaced by the most recent preceding $V_{sw}$ value available. {The magnitude $B$ of the Interplanetary Magnetic Field (IMF) was also taken into account, as well as the IMF clock angle $\theta$ in the Geocentric Solar Magnetospheric frame.}

{First, we considered the well-known solar wind-magnetosphere coupling parameter $V_{sw}^2B\sin^4(\theta/2)$ derived by \cite{Perreault78} and \cite{Vasyliunas82}. A high correlation $R=0.80$ was found between $Int(ap)$ and $Int(V_{sw}^2B\sin^4(\theta/2))$, the time-integrated $V_{sw}^2B\sin^4(\theta/2)$ over the same periods as $Int(ap)$, with a best least squares fit $Int(ap)=[Int(V_{sw}^2B\sin^4(\theta/2))]^{0.882}/574$. Therefore, this solar wind-magnetosphere coupling parameter largely controls the strength of $Int(ap)$ events. It is worth noting, however, that a slightly higher correlation ($R=0.87$) was found between $Int(ap)$ and $Int(V_{sw}^2 B)$ (see black points in Figure \ref{fig03}b), with a best fit $Int(ap)=[Int(V_{sw}^2B)]^{0.844}/3492$. These results are not unexpected in light of (i) the best correlation between $ap$ and $V_{sw}^nB$ found by \cite{Lockwood2013} for $n\simeq2$ (see also \cite{Petrukovich07}) and (ii) the higher degree of correlation for averaging timescales $>12$ hours between the $Am$ geomagnetic index and the coupling function $V_{sw}^2B$, as compared with the coupling function $V_{sw}^2B\sin^4(\theta/2)$ \cite{Lockwood2013}.}

{The dependence of $Int(ap)$ on solar wind-magnetosphere coupling parameters raised to powers $<1$ is consistent with at least a weak saturation of $Int(ap)$ starting to occur above 1000 nT$\cdot$hr inside the magnetosphere. However, the stronger saturation taking place above 5000 nT$\cdot$hr (see section 2.2) cannot be checked this way, due to the rarity of such very strong events in 2000-2017.}

{Correlations between $Int(ap)$ and the time-integrated $B$ or North-South IMF component $B_z$ are weaker, sensibly smaller than the correlation between $IntDst$ and $Int(|B_z|)$ found by \cite{Mourenas18}. As \cite{Wang2015} have noticed good correlations between the 3-hour $Kp$ index and peak values of some solar wind parameters during the preceding hours, we further examined different variants of this type and found a significant correlation ($R=0.64$, not shown) between $Int(ap)$ and the peak value over each $Int(ap)>1000$ nT$\cdot$hr event of the product of $V_{sw}(t_0)^2B(t_0)$ by the maximum value of $V_{sw}(t)^2B(t)$ over the 12 hours following $t_0$.}

{Finally, we explored possible correlations with the $AL$ index. The $AL$ index is intended to measure the auroral westward electrojet and, therefore, provides an estimate of the strength of energy loading and unloading processes occurring in the tail of the magnetosphere \cite{Davis1966, Gabrielse18}. Figure \ref{fig03}b demonstrates that $Int(ap)$ increases fast with $IntAL$, the time-integrated absolute value of $AL$ over $Int(ap)$ periods, since the best least squares fit to the data $Int(ap)=IntAL^{1.385}/281$ (green line) has a high correlation coefficient $R=0.79$. Therefore, the strength of the considered significant $Int(ap)$ events is also largely controlled by the energy loading and unloading processes occurring in the tail. The almost exponential increase, up to high levels, of $Int(ap)$ as a function of $IntAL$ further implies that the mechanisms responsible for the saturation of $Int(ap)$ should probably take place in the magnetotail or, earlier, at the solar wind-magnetosphere interface.  }

\subsection{{A preliminary scheme for predicting the strength of extreme $Int(ap)$ events}}

{Various recent studies \cite{Iucci2005, Ferguson2015, Kim2015, Borovsky17} have demonstrated a significant correlation between extended periods of high $ap$ and high fluxes of MeV electrons, as well as satellite anomalies. Therefore,} forecasting extreme $Int(ap)$ events could be useful to prevent adverse space weather effects on satellites by allowing, for instance, to momentarily shut them down during the few most risky periods \cite{HorneSW13, Schrijver2015}. {Accordingly, we tried to devise a possible prediction scheme of the strength of $Int(ap)$ events.} Assuming that past variations of $V_{sw}$, IMF magnitude $B$, and $ap$ can be used to provide a good probabilistic estimate of future variations \cite{McPherron2004, Wang2015, Owens2017} and building on {statistical results discussed in section 2.3,} we attempted to predict the 47 extreme events with $Int(ap)>2310$ nT$\cdot$hr that took place during 2000-2017.

{We reached an 83 \% probability of correct prediction (also called true positive rate) for these events when using the following three simultaneous thresholds: $\int_{t_0}^{t_0+12\,h}ap(t)dt>750$ nT$\cdot$hr, $V_{sw}(t_0)^2B(t_0) \times \max(V_{sw}(t)^2B(t))> 2.75\cdot10^{13}$ km$^4$nT$^2/$s$^4$ ($\max(x)$ being evaluated over $t_0\leq t\leq t_0+12$ h), and $ap(t_0+$11 h)$\geq 27$ nT. The corresponding probability of false alarms was $0.13$\%. Note, however, that this proposed predictor does not really forecast strong events before they occur: it only provides a warning at $t=t_0+$12 hours that an extreme event with $Int(ap)>2310$ nT$\cdot$hr has already started to occur, making use of solar wind and $ap$ measurements from the past 12 hours. Indeed, all such extreme events in 2000-2017} were predicted 8 to 72 hours ($\sim 25$ hours on average) before their end time. This is actually a key point for providing useful forecasts of the impact of such extreme events on satellites, because relativistic electrons fluxes usually reach their highest level near the end of high $ap$ periods \cite{Meredith2003, Thorne13:nature, Li2015, Tang17, Murphy2018}. {Combining the above predictor with other predictors of $ap$, $V_{sw}$, and IMF $B$ with several hours of lead time (e.g., \cite{Wing2005, Wang2015, Owens2017}) might enable one to further increase the lead time of the predictions, or to suppress false alarms.}

{However, we caution that the above-proposed prediction scheme should only be considered as a preliminary base for the development a truly reliable predictor. Indeed, it suffers from several important limitations that would need to be overcome before any practical use: (i) it was only tested over a limited 2000-2017 period, which includes a period of particularly weak geomagnetic activity in 2008-2017; (ii) some modulation with solar cycle should probably be taken into account; (iii) the degree of correlation between extreme $Int(ap)$ events and satellite anomalies has not yet been assessed; and (iv) the rate of false alarms still remains non-negligible. Improvements of the proposed prediction scheme along these lines are left for future work. }

\subsection{Return levels of extreme $Int(ap)$ events {and comparison with $Int(aa)$ events}}

It is useful for risk assessment to consider the $n$-year return level $Int(ap)_n$ of an extreme event -- the $Int(ap)$ level expected to be exceeded once every $n$ years. For extreme $Int(ap)>5200$ nT$\cdot$hr events in 1958-2007 having a Generalized Pareto Distribution, it can be estimated as $Int(ap)_n\simeq \min[Int(ap)]+(\sigma/\xi)[(41 n/50)^\xi-1]$, with $\xi\sim -0.33$ and $\sigma\sim 3060$ \cite{Coles2001}. When considering the power-law distribution ML fit with upper cutoff obtained in section 2.2, it is given by $Int(ap)_n = [50/(236n)(\min[Int(ap)]^{-\alpha+1} -\max[Int(ap)]^{-\alpha+1}) +\max[Int(ap)]^{-\alpha+1}]^{1/(-\alpha+1)}$, with $\alpha=2.85$, $\min[Int(ap)]\simeq 2310$ nT$\cdot$hr, and $\max[Int(ap)]\simeq 13000$ nT$\cdot$hr. {A solar cycle modulation of the number of extreme $Int(ap)$ events by a factor $S_C\sim(\max[S_{n,y}]/210)^2$ (see section 2.3) can easily be taken into account in the above $Int(ap)_n$ formulas by multiplying $n$ by $S_C$.}

Figure \ref{fig04}a shows that the GPD and power-law with upper cutoff distribution fits yield very similar return levels, increasing with the number $n$ of years until they reach a similar upper limit $\max[Int(ap)]\simeq 13000-14000$ nT$\cdot$hr. Modelled return levels based on these fits are close to observed return levels over 1958-2007. All the observed return levels are nearly comprised between the modelled return level based on the GPD fit and the maximum (for a 95\% confidence interval) modelled return level based on the power-law fit. {Even the highest possible return level (for a 95\% confidence interval) of the GPD fit remains smaller than $\sim14000$ nT$\cdot$hr over 100 years, but its upper limit on much longer timescales is much larger $\sim800000$ nT$\cdot$hr.} These results show that $Int(ap)$ events larger than $14000$ nT$\cdot$hr are very unlikely to be observed in the next 50-100 years without an important change in the solar wind statistical behavior. Moreover, the largest event of the 1958-2017 period, which occurred in October 2003 and reached 12700 nT$\cdot$hr, may be considered as a typical 1 in $100\pm40$ years event.

\begin{figure*}
\includegraphics[width=36pc]{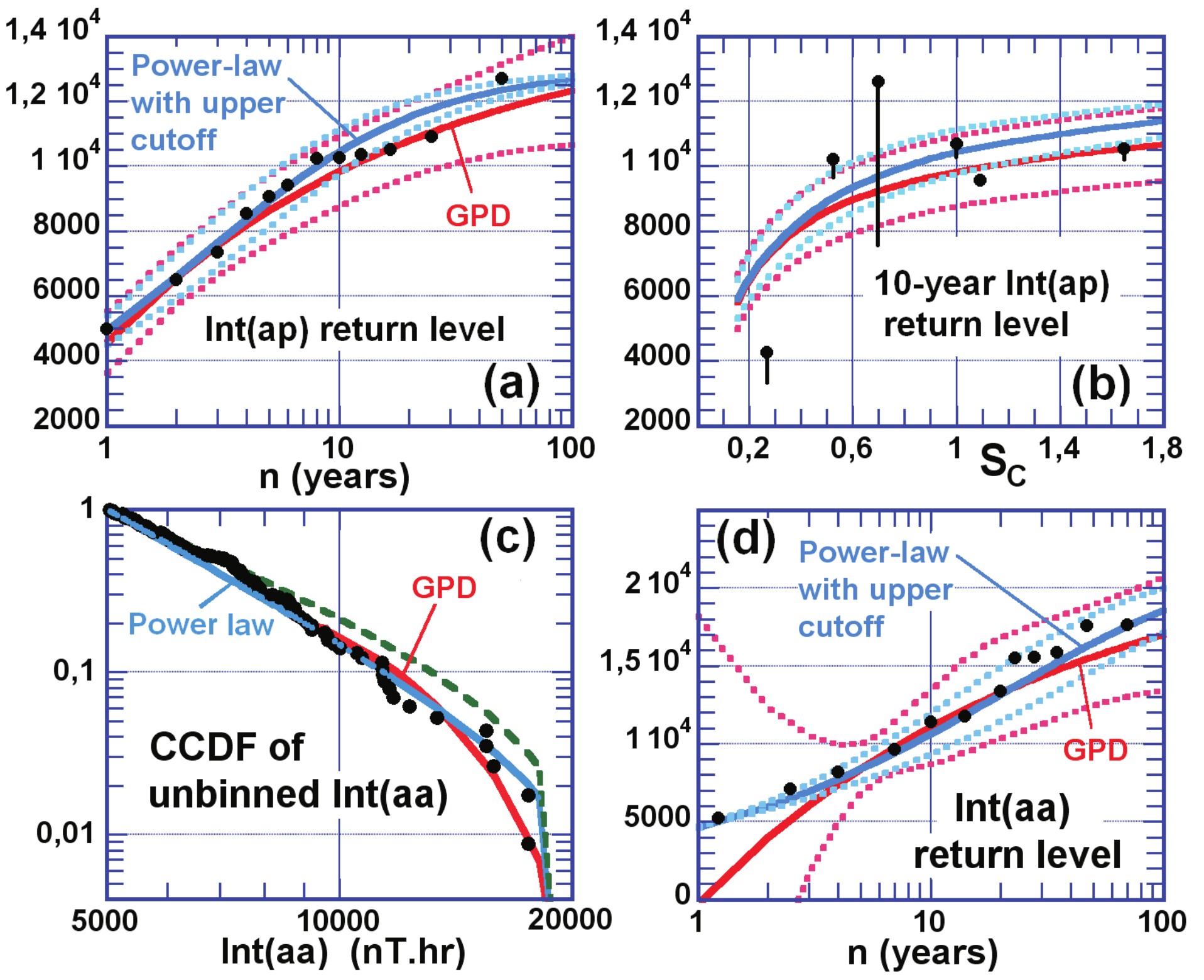}
\end{figure*}
\begin{figure*}
\caption{(a) Modelled return levels of extreme $Int(ap)$ events as a function of the number $n$ of years, calculated based on the GPD ML fit (solid red curve) and on the power-law ML fit with upper cutoff (solid blue curve), with 95\% confidence intervals calculated via the delta method \cite{Coles2001} indicated by light dotted curves (the mean $S_C=1$ value is used). Observed return levels in 1958-2007 are also shown (black points). (b) Modelled 10-year return levels of extreme $Int(ap)$ events over approximately one solar cycle, based on GPD (red) and power-law (blue) ML fits (with their 95\% confidence intervals assuming no $S_C$ uncertainty), as a function of the factor $S_C$ that takes approximately into account the variation of the total number of extreme events with maximum yearly mean sunspot number during a solar cycle. Observed return levels during the past 6 solar cycles are shown by black points, with uncertainty bars corresponding to the distance to the second strongest 10-year event during each cycle. {(c) CCDF of extreme $Int(aa)>5000$ nT$\cdot$hr events in 1870-2010 (black circles) and corresponding ML fits in the form of a power-law with upper-cutoff (blue) or a Generalized Pareto Distribution (in red for $Int(ap)>8900$ nT$\cdot$hr). A power-law fit with the same upper cutoff but $\alpha=3$ like the $Int(ap)$ distribution is also plotted (green dashed curve). (d) $n$-years return levels of extreme $Int(aa)$ events estimated based on GPD (red) and power-law (blue) ML fits. Observed return levels in 1870-2010 are also shown (black points).} }
\label{fig04}
\end{figure*}

Let us caution that the above results are based on the statistical distribution of $Int(ap)$ events over 1958-2007. Modelled return levels displayed in Figure \ref{fig04}a implicitly assume that this 50-year time-averaged distribution is well representative of the 1-year, 10-year, or 100-year distributions. However, we have seen in section 2.3 that the number of extreme events during each solar cycle is roughly modulated by a factor $S_C$ depending on the maximum yearly mean sunspot number during this cycle. The results in Figure \ref{fig04}a therefore assume that $S_C=1$, which is the average value of $S_C$ over 1958-2007. But when trying to estimate return levels over a given solar cycle, a significant modulation by $S_C$ ought to be taken into account. The variation of the 10-year $Int(ap)_{10}$ modelled return levels with $S_C$ is displayed in Figure \ref{fig04}b over the range $0.15\leq S_C\leq 1.8$ encompassing all values taken by $S_C$ during the past 317 years. The modelled return level merely varies by 20\% over $0.6<S_C<1.8$, but it decreases quickly as $S_C$ decreases below $0.6$. Solar cycles characterized by a particularly low sunspot activity (as the present cycle or during a Dalton or Maunder minimum) can lead to 30\% to 50\% smaller 10-year return levels than the estimate obtained with $S_C=1$.

Comparisons between modelled 10-year return levels based on the GPD and power-law fits and observed 10-year return levels during the past 6 solar cycles demonstrate a reasonable agreement in Figure \ref{fig04}b {(the relative discrepancy remains always less than 42\%).} However, the very low 10-year return level observed during the present (not yet complete) solar cycle and the highest return level observed during the preceding cycle lie outside the 95\% confidence interval of the model, calculated with an assumed null $S_C$ uncertainty. {This means that the actual uncertainty on the solar cycle modulation factor of 10-year return levels is probably larger than 30\%.}

{When considering future events, however, any use of probabilistic estimates derived from available datasets requires to assume a quasi-stationarity of the distribution in the near future. This assumption will be partly tested below through comparison of the 1958-2007 dataset of $Int(ap)$ events with the much larger 1870-2010 dataset of $Int(aa)$ events.} The $aa$ index (in nT units), which roughly mimics $ap$ on long timescales, is derived from 3-hour measurements at two antipodal stations in England and Australia, normalized to geomagnetic latitudes $\pm 50^\circ$ \cite{Mayaud80}. The integration threshold for $Int(aa)$ events was fixed at $aa\geq 32$ nT to roughly match the $ap\geq 22$ nT threshold used for $Int(ap)$ (it is also the highest $aa$ threshold that does not lead to a splitting of some large $Int(ap)$ events into separate, smaller $Int(aa)$ events).

{A GPD ML fit to the extreme $Int(aa)$ distribution was obtained by the same EVT method as before, giving $\xi\sim -0.28 \pm 0.6$ and $\sigma\sim 4100 \pm3000$ for $Int(aa)>8900$ nT$\cdot$hr. Although the 95\% confidence interval for $\xi$ is large and encompasses some positive values, $\xi$ still remains negative within a 64\% confidence interval, implying the probable existence of an upper limit. Based on the optimal GPD parameters, this upper limit is $\sim 22000-25000$ nT$\cdot$hr (see Figure \ref{fig04}c). This upper limit is only slightly larger than the highest observed level $Int(aa)\simeq 18000$ nT$\cdot$hr, reached during both the May 1921 \cite{Cliver2013} and October 2003 superstorms,  which represent 1 in 50-120 years events (see return levels in Figure \ref{fig04}d). This lends further credence to an upper limit on $Int(ap)$ taken as slightly larger than the level of the October 2003 event.}

{A power-law ML fit $P_y\simeq C/Int(aa)^\alpha$ with upper cutoff $\max[Int(aa)]=22000$ nT$\cdot$hr to the yearly distribution of $Int(aa)>5000$ nT$\cdot$hr events gives $\alpha\sim 3.59 \pm0.59$ and $C\sim 8\cdot10^{9}$ (see Figure \ref{fig04}c). The maximum KS distance $D\simeq 0.1$ of this fit to the CCDF of the data corresponds to $p=0.32$, showing that the power-law distribution hypothesis cannot be confidently rejected.} A second power-law fit with $\alpha\sim 3$ (as for the $Int(ap)$ distribution) and $\max[Int(aa)]=1.7\times\max[Int(ap)]$ {is plotted in Figure \ref{fig04}c (green dashed curve) and remains close to the data. It shows that the probability distributions of extreme $Int(ap)$ and $Int(aa)$ events can be fitted by similar power-law functions, with similar exponents,} and also similar upper cutoffs once the actual $\approx 1.7$ ratio of $ap$ over $aa$ units \cite{Lockwood2018} is taken into account. {Since the $Int(aa)$ dataset covers a much longer 141-year period, the hypothesis of a quasi-stationarity of the $Int(ap)$ distribution over the long term appears reasonable -- as long as important changes in the sun statistical behavior are not expected to occur. }

\subsection{Relationships between $Int(ap)$ and {particle fluxes in the magnetotail and outer radiation belt}}

{In section 2.3, the strength of $Int(ap)$ events was shown to be largely controlled by the solar wind-magnetosphere coupling, mainly through energy loading and unloading processes occurring in the tail. This expected strong impact of substorm-related injections is worth checking with satellite measurements of incoming energetic particles in the near-Earth magnetotail and 150 keV electron injections at geosynchronous orbit. Moreover, recent works have found significant correlations between time-integrated measures of $Kp$ and MeV electron flux \cite{Borovsky17, Borovsky17b, Kim2015} as well as satellite anomalies \cite{Ferguson2015, Iucci2005} in the outer radiation belt. The present study of $Int(ap)$ events was actually undertaken based on these correlations. Therefore, this study would not be complete without at least a brief investigation of MeV electron flux levels reached after $Int(ap)$ events in the heart of the outer radiation belt, as a function of the strength of these events.}

{Significant disturbances are often accompanied, during their initial phase, by a strong solar wind dynamic pressure impulse that compresses the magnetosphere and leads to a dropout of 100 keV to multi-MeV electrons via magnetopause shadowing (e.g., \cite{Shprits06, Boynton16, Boynton17, Murphy2018} and references therein) before electron fluxes eventually recover and even increase. As a practical consequence, such dropouts disconnect the final and initial states of the radiation belt, leading to a very weak dependence of the final (after a storm) electron flux on its initial level (e.g., \cite{Murphy2018, Reeves2003}). This provides us with an opportunity to better assess the impact of the sole $Int(ap)$ parameter on MeV electron flux, by only considering events with $Int(ap)>1000$ nT$\cdot$hr in 2010-2017 such that a solar wind dynamic pressure impulse reached $>10$ nPa during the preceding day and caused a dropout of 1.8 MeV electron flux at $L^*\sim4.5$. Events with smaller dynamic pressure impulse (and potentially no flux dropout) that could correspond to important fluxes independently of the $Int(ap)$ effect, were therefore discarded. We also discarded events occurring less than two days from other events, finally keeping 12 moderately intense events with $Int(ap)\sim 1000-2000$ nT$\cdot$hr and 6 strong ones with $Int(ap)\sim 2300-4200$ nT$\cdot$hr.}

{First, we investigated only $Int(ap)$ events with simultaneously available measurements from Time History of Events and Macroscale Interactions during Substorms (THEMIS) satellites at $L=9-12$ in the near-Earth magnetotail \cite{Angelopoulos08}. For each event, we selected all subintervals characterized by enhanced transport/injection of magnetotail particles (earthward plasma velocity exceeding $100$ km/s) and used combined measurements of the Electrostatic Analyzer and Solid State Telescope onboard THEMIS D \cite{Angelopoulos08, McFadden08:THEMIS}. The typical total duration of all subintervals in one event was several hours ($\sim 10-20$\% of total event duration). To characterize the efficiency/power of injections from the magnetotail into the inner magnetosphere, we calculated the total inward transported flux of particles over each subinterval, integrating ion and electron earthward fluxes over energy and time. The resulting energy density $J_{v_x>0}$ is an average energy per element of transverse $(z,y)$ injection area.} The total injection area in the $(z,y)$ plane can be estimated based on statistical multispacecraft measurements \cite{Nakamura04}. {Finally, the distribution of injections collected during each event was renormalized to their MLT-averaged occurrence rate \cite{Gabrielse14}. The obtained parameter $J_{v_x>0}$ can be used to characterize injection input to the inner magnetosphere (since there is a clear correlation between the $J_{v_x>0}$ distribution and the $IntDst$ parameter, see \cite{Mourenas18}). The CCDF of $J_{v_x>0}$ displayed in Figure \ref{fig05}a shows that the time-integrated earthward particle energy flux above $4\times 10^8$ keV$/$cm$^2$ is significantly larger during strong ($Int(ap)>2300$ nT$\cdot$hr) events than during weaker ($Int(ap)<1600$ nT$\cdot$hr) events. Next, we used GOES 15 electron flux  measurements at geosynchronous orbit to evaluate the distribution of the 10-min maximum 150 keV electron flux during typically one hour of rapid substorm injections for each event. Figure \ref{fig05}b shows that the CCDF of the 10-min maximum flux of 150 keV electrons during injections has also a harder spectrum during strong events. Thus, harder energy spectra of substorm-injected particles at $L=9-12$ and $L\sim6.6$ correspond in general to stronger $Int(ap)$ events. }

\begin{figure*}
\centering
\includegraphics[width=36pc]{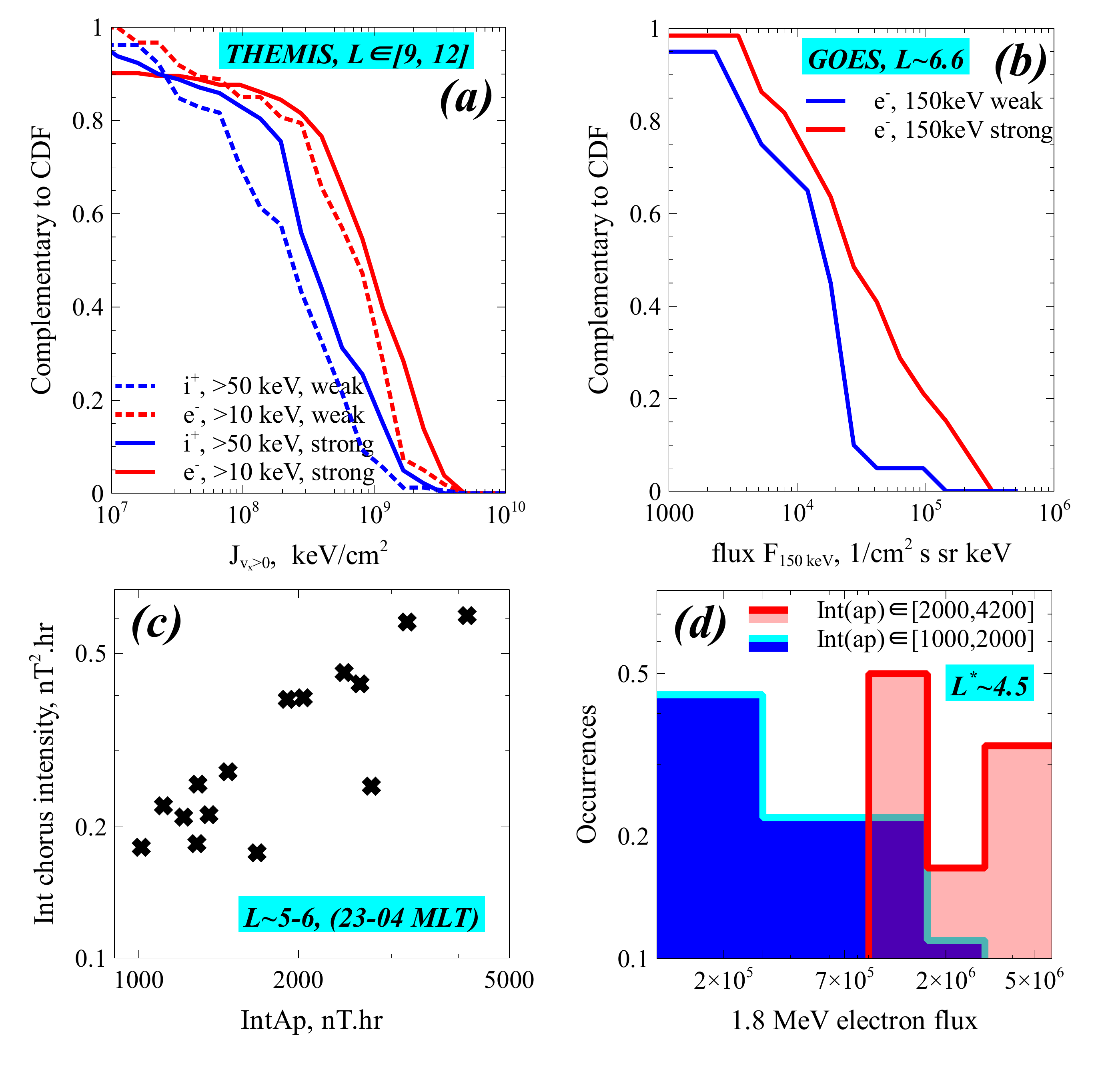}
\end{figure*}
\begin{figure*}
\caption{(a) CCDF of time integrated earthward energy fluxes of $>50$ keV ions and $>10$ keV electrons in the magnetotail from MLT-integrated THEMIS measurements for weak ($Int(ap)<1600$ nT$\cdot$hr) and strong ($Int(ap)>2300$ nT$\cdot$hr) events in 2010-2017. (b) CCDF of 150 keV 10-min maximum flux of injections measured by GOES at $L=6.6$ during the same events. (c) Time-integrated lower-band chorus wave intensity at $L\sim 5$ from a statistical chorus model \cite{Agapitov18} during the 15 selected events in 2013-2017. {(d) Distribution of daily mean 1.8 MeV electron flux (in e/cm$^2$/s/sr/MeV) measured by the Van Allen Probes at $L^*=4.25-4.75$ just after the same 15 events, for weak (blue) and strong (red) events. The spacecraft $L^*$ position was determined using the TS04 \cite{Tsyganenko05} external magnetic field model and the International Geomagnetic Reference Field internal field model.} }
\label{fig05}
\end{figure*}

{Figure \ref{fig05}d further shows the distribution of the daily mean flux of 1.8 MeV electrons measured by the Van Allen Probes \cite{Baker13} at adiabatically invariant $L$-shell $L^*=4.25-4.75$ just after the end of the 15 selected events that occurred in 2013-2017, plotted separately for weak/moderate events ($1000-2000$ nT$\cdot$hr, in blue) and strong events ($2000-4200$ nT$\cdot$hr, in red). Figure \ref{fig05}d demonstrates that strong events generally lead to higher 1.8 MeV electron fluxes than weak/moderate events. For instance, a high daily flux $>9\times10^5$ e/cm$^2$/s/sr/MeV of 1.8 MeV electrons is reached after 100\% of the strong events, but only after 33\% of the weak/moderate events. Very high fluxes $\sim 5\times10^6$ e/cm$^2$/s/sr/MeV are only reached after strong events.}

{Elevated MeV electron fluxes often result from the cumulative effects of generated chorus and ULF waves (e.g., \cite{Ma18}). Powerful chorus waves generated by strong injections of anisotropic $10-30$ keV electrons can efficiently accelerate incoming $100-300$ keV electrons up to MeVs at $L\sim 4-6$ near the end of high-$Kp$ periods \cite{Artemyev15:nature, Boyd18, Li2015, Meredith2003, Mourenas18b, Tang17, Thorne13:nature, Zhao17}. The time-integrated intensity of lower-band chorus waves at $L\sim 5$ in the 23-04 MLT sector, estimated during the same 15 events based on the statistical chorus model from \cite{Agapitov18}, increases with $Int(ap)$ in Figure \ref{fig05}c.  Therefore, this acceleration mechanism can be very efficient during strong events. Intense and persistent ULF waves generated by injected ions \cite{Ozeke08} or by prolonged solar wind variations (e.g., \cite{Zong2017}) can also diffuse electrons radially inward and accelerate them \cite{Ozeke08, Ozeke14, Zong2017}. A significant correlation was found between time-integrated ULF wave intensity and MeV electron flux at $L=6.6$ \cite{Kozyreva07}. Since ULF wave intensity is known to statistically increase with $Kp$ \cite{Ozeke14}, the time-integrated ULF wave intensity is expected to be larger during strong $Int(ap)$ events.}

{However, we caution that the above results are based on a limited dataset of 15 significant $Int(ap)$ events during a period of relatively weak geomagnetic activity in 2013-2017. A larger statistics of significant ($>1000$ nT$\cdot$hr) events would be needed to draw more definitive conclusions. In addition, other parameters can affect the level of MeV electron flux relatively independently of $Kp/ap$, such as solar wind velocity and density \cite{Borovsky17}, or even localized electric fields and the strength of dipolarizations \cite{Gabrielse14, Ganushkina2017}. The presence of such additional parameters and physical processes probably explains the variability of MeV electron fluxes for a given range of $Int(ap)$ in Figure \ref{fig05}d and should not be overlooked.}

{Still, it is worth noting that the high daily flux $\sim(1-6)\times10^6$ e/cm$^2$/s/sr/MeV of 1.8 MeV electrons reached after strong events is expected to remain elevated over days to weeks after the end of each event, because dropouts of 1.8 MeV electron flux occur only every $\approx 20-100$ days at $L\sim4.5$ \cite{Boynton17}, while the decay of 1.8 MeV electron flux inside the plasmasphere via precipitation by hiss waves into the atmosphere is slow during low geomagnetic activity at such $L$-shells \cite{Mourenas17}. We checked that daily fluxes of 1.8 MeV electrons at $L^*\sim4.5$ indeed remained above $5\times10^5$ e/cm$^2$/s/sr/MeV during 9 to 14 days after each strong event. }

\section{Conclusions}

In this paper, we have provided a large statistical study of extreme time-integrated $ap$ (denoted $Int(ap)$) events of prolonged and continuously elevated geomagnetic activity during the period 1958-2007. Such extreme $Int(ap)$ events are expected to correspond to high fluxes of MeV (and $30-150$ keV) electrons in the outer radiation belt, and to higher rates of satellite anomalies at $L=4.0-6.6$. These events occur during both strong and weak geomagnetic storms, corresponding to both high and moderate time-integrated $Dst$, making the $Int(ap)$ distribution significantly different from the previously obtained $IntDst$ distribution \cite{Mourenas18}.

The tail of the distribution at $Int(ap)>2300$ nT$\cdot$hr can be well fitted by a power-law function with a sharp upper cutoff, just like the distribution of extreme $IntDst>1500$ nT$\cdot$hr events \cite{Mourenas18}. Roughly half of the events belong simultaneously to both populations above their respective thresholds -- much more than below those thresholds. The consistent emergence of such power-law distributions suggests that the solar wind-magnetosphere-ionosphere system, under sufficiently strong and persistent driving defined by $Int(ap)>2300$ nT$\cdot$hr or $IntDst>1500$ nT$\cdot$hr, may reach some sort of self-organized critical configuration (e.g., \cite{Angelopoulos1999, Aschwanden16}).

In general, such a marginally stable critical configuration constitutes an attractor for the system dynamics and it is naturally reached through a self-organizing process \cite{Aschwanden16}. Could it be the same for the magnetospheric system? The observed sharp upper cutoff on $Int(ap)$ does provide evidence for the possible presence of a self-limiting mechanism, which could take the form of various large-scale physical processes known to show up during sufficiently strong and sustained driving, such as: (i) an increased deflection of solar wind flow \cite{Lopez10} hindering magnetic reconnection and therefore reducing injections, (ii) an increased auroral precipitation of substorm-injected particles leading to more ionospheric O$^+$ outflow that may in turn decrease the nightside reconnection rate or the number of reconnection sites in the tail \cite{Liu2013, Schillings17, Varney2016, Welling2016}, or (iii) an increased and deeper wave-mediated energy transfer from injected to relativistic electrons \cite{Agapitov18, Li2015, Meredith2003, Mourenas18b}. Such a self-saturated, nearly stable configuration could represent a natural attractor for the strongly driven solar wind-magnetosphere-ionosphere system and correspond to more long range internal correlations, potentially explaining the power-law distribution of extreme $Int(ap)$ events. Various observations and studies have hinted at the possible existence of a self-organized state of the magnetosphere-ionosphere system during strong driving with sustained substorm injections \cite{Angelopoulos1999, Klimas2000, Uritsky2001, Valdivia2013, Varney2016, Welling2016}. However, further work will be needed to investigate these conjectures in more details.

Based on the accuracy and similarity above $Int(ap)=5000$ nT$\cdot$hr of the power-law distribution fit with upper cutoff and the Generalized Pareto Distribution fit with negative (over a 95\% confidence interval) shape parameter, the existence of an upper limit $\max[Int(ap)]\simeq 13000-14000$ nT$\cdot$hr on the strength of $Int(ap)$ events is likely. The similarity of the 1870-2010 $Int(aa)$ distribution to the 1958-2007 $Int(ap)$ distribution further supports the applicability of the obtained $Int(ap)$ distribution fits and upper limits over 150-year periods, provided that the modulation of the probability of extreme events with solar cycle is taken into account over periods smaller than 20-30 years. Based on the Maximum Likelihood GPD fit, the 1 in 10 years and 1 in 100 years levels of extreme $Int(ap)$ events are $9700 \pm 1000$ and $12350 \pm 1700$ nT$\cdot$hr, respectively, the latter being close to the expected upper limit. The Maximum Likelihood power-law fit with upper cutoff gives similar 1 in 10 years and 1 in 100 years levels of $10400 \pm 700$ and $12700 \pm 130$ nT$\cdot$hr, respectively. The October 2003 event with $Int(ap)=12700$ nT$\cdot$hr is representative of a 1 in $100\pm40$ years event. During the earlier 1870-1957 period, only the May 1921 event reached an $Int(aa)$ level similar to the October 2003 event. As for the famous Carrington 1859 superstorm, its $IntDst$ level was apparently smaller than for the October 2003 event \cite{Mourenas18}, which could also imply a smaller $Int(ap)$. In contrast, during periods of particularly low solar activity such as a Dalton minimum, the 10-year return level is expected to decrease by 30\%-50\% at least as compared to its time-averaged level given above. Comparisons with future observations should allow to test the accuracy of these probabilistic forecasts, while adding more recent data to the 1958-2007 $Int(ap)$ dataset should allow to improve the present models.

{An analysis of correlations with solar wind and geomagnetic activity parameters has shown that the strength of $Int(ap)$ events is largely controlled by the solar wind-magnetosphere coupling, mainly via energy loading and unloading processes occurring in the tail. Considering a limited dataset of significant events in 2010-2017, we further found that strong events above $2000-2300$ nT$\cdot$hr indeed correspond in general to harder and more sustained 10-200 keV ion and electron injections coming from the magnetotail into the outer radiation belt. Moreover, such strong events can lead to high daily fluxes $>9\times10^5$ e/cm$^2$/s/sr/MeV of 1.8 MeV electrons near $L^*\sim4.5$, apparently higher in general than after weaker events. Such elevated MeV electron fluxes probably result from the cumulative effects of chorus and ULF waves amplified during the event \cite{Ma18}.  }

{Finally, several previous studies have provided experimental evidence for the existence of an upper limit on MeV electron flux \cite{OBrien07, Meredith17}. However, they considered less than 20 years of data. Taken together, the presence of an upper limit on $Int(ap)$ and the correlation of MeV electron flux with prolonged periods of high $ap$ \cite{Borovsky17, Ferguson2015, Kim2015} add more weight to this hypothesis. }

\section{Acknowledgements}
We used 1958-2017 data of $ap$, IMF, $V_{sw}$, $Dst$, $AL$ from the World Data Center in Kyoto (http://wdc.kugi.kyoto-u.ac.jp/kp/index.html) and OMNIweb (http://omniweb.gsfc.nasa.gov), the latest International Sunspot Number from WDC-SILSO (http://www.sidc.be/silso/), 1958-2007 $Dst$ data from the U.S. Geological Survey (https://geomag.usgs.gov/), and $aa$ data from ISGI Collaborating Institutes (http://isgi.unistra.fr). X.J.Z. acknowledges support from RBSP-EMFISIS funding 443956-TH-81074 under NASA's prime contract NNN06AA01C. A.V.A., X.J.Z., and V.A. acknowledges NASA contract NAS5-02099 for using THEMIS data from http://themis.ssl.berkeley.edu/. We thank C.W. Carlson and J.P. McFadden for the use of ESA data, D.E. Larson and R.P. Lin for the use of SST data, and K.H. Glassmeier, U. Auster, and W. Baumjohann for the use of FGM data provided under the lead of the Technical University of Braunschweig and with financial support through the German Ministry for Economy and Technology and the German Aerospace Center (DLR) under contract 50OC0302. We gratefully acknowledge GOES teams for data available at https://www.ngdc.noaa.gov/, and Van Allen Probes teams for the data available at http://www.rbsp-ect.lanl.gov/data-pub/rbspa/. D.M. would also like to thank L. Lyons for suggesting to examine the $AL$ index.

\bibliographystyle{agu}

\end{document}